\documentclass[a4paper]{article}

\usepackage{INTERSPEECH2021}
\usepackage{hyperref}
\usepackage{algorithm}      
\usepackage[noend]{algpseudocode} 

\newcommand{\algrule}[1][.2pt]{\par\vskip.1\baselineskip\hrule height #1\par\vskip.1\baselineskip}
\algdef{SE}[DOWHILE]{Do}{doWhile}{\algorithmicdo}[1]{\algorithmicwhile\ #1}%

\title{Self-paced ensemble learning for speech and audio classification}
\name{Nicolae-C\u{a}t\u{a}lin Ristea$^1$, Radu Tudor Ionescu$^2$}

\address{
  $^1$University Politehnica of Bucharest, Romania\\
  $^2$University of Bucharest, Romania}
\email{r.catalin196@yahoo.ro, raducu.ionescu@gmail.com}

\begin{document}

\maketitle

\begin{abstract}
Combining multiple machine learning models into an ensemble is known to provide superior performance levels compared to the individual components forming the ensemble. This is because models can complement each other in taking better decisions. Instead of just combining the models, we propose a self-paced ensemble learning scheme in which models learn from each other over several iterations. During the self-paced learning process based on pseudo-labeling, in addition to improving the individual models, our ensemble also gains knowledge about the target domain. To demonstrate the generality of our self-paced ensemble learning (SPEL) scheme, we conduct experiments on three audio tasks. Our empirical results indicate that SPEL significantly outperforms the baseline ensemble models. We also show that applying self-paced learning on individual models is less effective, illustrating the idea that models in the ensemble actually learn from each other.
\end{abstract}

\setlength{\abovedisplayskip}{3pt}
\setlength{\belowdisplayskip}{3pt}

\vspace{-0.1cm}
\section{Introduction}

State-of-the-art machine learning models are trained using supervision in the form of labeled data samples provided by an oracle (a domain expert or an ordinary human that knows how to label the samples). However, effective human learners have the ability to guide themselves through the learning activities, considering increasingly difficult concepts at their own pace, without the help of an oracle (a teacher). Indeed, most of the times, students choose what, how, when and how long to study, implicitly using a self-paced curriculum. These self-regulated aspects of learning influence the future results of the learner. This cognitive learning paradigm inspired researchers to study and propose machine learning models based on self-paced learning (SPL) \cite{kumar2010self, jiang2015self}. We underline that SPL can be seen as a particular case of curriculum learning \cite{Bengio-ICML-2009,Soviany-A-2021}, in which the difficulty of the examples is estimated by the model itself, hence the self-pacing. A variety of SPL schemes have been designed for different computer vision \cite{jiang2015self, lin2020pixel} and other pattern recognition tasks \cite{zheng2020unsupervised, zhou2020self}, with demonstrated improvements over the standard supervised learning paradigm in specific scenarios. In signal processing, there are only a handful of approaches based on SPL \cite{shang2018self}. 

In the most straightforward SPL approach, the model uses its own confidence scores to determine what are the next examples to learn. However, if the examples come from a different (target) domain and their ground-truth labels are not available, the model can only use the labels predicted by itself (pseudo-labels). Therefore, in the context of unsupervised domain adaptation, we encounter a paradox: there is little new information to be learned from pseudo-labeled examples that are selected by the model with high confidence. To overcome the paradox, some methods \cite{Soviany-WACV-2020,Soviany-CVIU-2021} use an external difficulty predictor. In this work, we conjecture that useful information can be learned simply by training multiple models together. To this end, we propose to jointly train an ensemble of models that learn from each other through SPL. More precisely, an individual model from the ensemble will benefit from the pseudo-labels and the confidence scores provided by the whole ensemble, thus being able to take useful information from the other models included in the ensemble. To the best of our knowledge, we are the first to propose self-paced ensemble learning (SPEL) in the audio domain.
By combining self-paced learning and ensemble learning into a single framework, there are two essential benefits: higher effectiveness and increased generality. To demonstrate the benefits of SPEL, we conduct experiments on three audio benchmark tasks, namely speech emotion recognition on CREMA-D \cite{Cao-TAC-2014}, speech mask detection on MSC \cite{Schuller-INTERSPEECH-2020} and bird and frog species detection on RCSAD \cite{RFCx}. Our empirical results indicate that SPEL provides superior results than the baseline ensembles and the state-of-the-art methods. In addition, we present ablation results on CREMA-D showing that employing self-paced learning on an individual model is less effective, indicating that models in an ensemble can effectively learn from each other.

\section{Related Work}

\noindent {\bf Ensemble learning.}
On the one hand, traditional machine learning models may fail to attain the desired performance level when dealing with high-dimensional, noisy or imbalanced data. The reason behind this is that such models belong to restricted hypothesis classes, thus not being able to discover complex patterns revealing the underlying structure of the data. On the other hand, deep neural networks can model such complex patterns when sufficient data is available, but such models lack stability, mainly due to the random initialization of parameters and the stochastic optimization process. In both cases, ensemble learning \cite{dong2020survey,Zhang-INTERSPEECH-2015} offers improved effectiveness and stability by integrating multiple models into a single framework. There are several ensemble learning methods in literature, ranging from majority voting \cite{raza2019improving} and Bayesian model averaging \cite{fersini2014sentiment} to bagging \cite{bryll2003attribute}, boosting \cite{jianqiang2016combing} and classifier stacking \cite{sikora2015modified}. As other ensemble learning methods applied in the audio domain \cite{elowsson2017predicting, zahid2015optimized}, we employ Bayesian model averaging in our method. Different from the related methods, we adapt our ensemble to the target domain using self-paced learning, attaining superior results.

\noindent {\bf Self-paced learning.}
The vast majority of SPL approaches proposed in literature are focused on computer vision topics, the technique being proved effective in many vision tasks \cite{sangineto2018self, zhou2018deep}. Sangineto et al.~\cite{sangineto2018self} proposed a new protocol based on the self-paced learning paradigm, where the main idea is to iteratively select a subset of images and bounding boxes that are the most reliable, and use them for training. The proposed method outperforms related weakly-supervised object detectors, attaining state-of-the-art results. Zhou et al. \cite{zhou2018deep} designed a model based on an SPL regime performing person re-identification, which usually suffers from noisy samples with background clutter and mutual occlusion, making it extremely difficult to distinguish different individuals across disjoint camera views. Their solution seems to overpass these issues, attaining superior performance in comparison with state-of-the-art approaches.

Different from these methods \cite{sangineto2018self, zhou2018deep}, we employ an algorithm that combines the advantages of ensemble learning and SPL, resulting in a more effective and stable approach.

\noindent {\bf Self-paced ensemble learning.}
Pi et al.~\cite{pi2016self} proposed a self-paced boost learning method that incorporates an adaptive easy-to-hard pace in the boosting process. The method is demonstrated on generic tasks using weak learners such as Logistic Regression or Support Vector Machines. Liu et al.~\cite{Liu-ICDE-2020} introduced a self-paced ensemble method to cope with the class imbalance problem. The method is based on combining self-paced learning with undersampling to generate a robust ensemble. As Pi et al.~\cite{pi2016self}, they demonstrated the utility of the proposed method on ensembles of weak learners. Since deep neural networks require large data sets during training, the undersampling approach of Liu et al.~\cite{Liu-ICDE-2020} is not suitable for deep models. In contrast, we propose an ensemble of deep models and apply SPL for unsupervised domain adaptation.

SPL has also been applied on clustering methods. For example, Zhou et al.~\cite{zhou2020self} proposed a novel self-paced clustering ensemble that gradually adds instances, from easy to difficult, into the ensemble learning. Their approach overcomes the existing problems of ensemble clustering methods, which usually exploit all data samples to learn a consensus clustering result, ignoring the adverse effects caused by difficult instances. Ghasedi et al.~\cite{ghasedi2019balanced} proposed a different clustering approach, in which a balanced self-paced learning algorithm improves the training process of deep generative adversarial clustering networks, tackling the drawbacks of training deep clustering models in an unsupervised manner. Unlike these methods~\cite{zhou2020self,ghasedi2019balanced}, we focus on supervised classification and detection problems.

Wang et al.~\cite{Wang-TMCCA-2020} used SPL in a very particular scenario, focusing on a Random Forest ensemble applied specifically to lung cancer prognosis. Different from Wang et al.~\cite{Wang-TMCCA-2020}, we consider ensembles composed of various deep neural architectures applied on a broad variety of audio tasks.

Unlike the aforementioned methods \cite{zhou2020self,pi2016self,Liu-ICDE-2020,ghasedi2019balanced,Wang-TMCCA-2020}, we employ an ensemble of deep neural networks that are $(i)$ trained on the source domain using standard supervision, then $(ii)$ fine-tuned on the target domain using self-paced learning with pseudo-labeling, leading to a more robust ensemble with superior results on multiple audio classification and detection tasks. To the best of our knowledge, there is no prior work based on self-paced ensemble learning in the audio domain.

\section{Method}


\noindent {\bf Data representation.}
We first transform each audio sample into a 2D representation in order to be able to employ state-of-the-art convolutional neural networks (CNNs) \cite{He-CVPR-2016,Krizhevsky-NIPS-2012} based on 2D convolutions. To this end, we compute the discrete Short Time Fourier Transform (STFT), as follows:
\begin{equation}\label{eq_stft}
STFT\{x[n]\}(m, k)=\!\!\sum_{n=-\infty}^{\infty}\!\! x[n] \cdot w[n-m R] \cdot e^{-j \frac{2 \pi}{N_x}k n},
\end{equation}
where $x[n]$ is the discrete input signal, $w[n]$ is a window function (in our approach, Hann), $N_x$ is the STFT length and $R$ is the hop (step) size \cite{Allen-IEEE-1977}. Afterwards, we compute the spectrograms as the squared magnitude of the STFT, convert the resulting values to a logarithmic scale (decibels) and normalize them to the interval $[-1,1]$, generating a single-channel grayscale image. Moreover, the frequency bins were mapped onto the Mel scale with 256 Mel bands.

\noindent {\bf Baseline model ensemble.}
As our learning approach, we consider an ensemble of residual neural networks \cite{He-CVPR-2016,zhang2020resnest} that are combined through Bayesian model averaging. We employ residual nets because such architectures are known to produce state-of-the-art results on a wide variety of tasks, mainly due to their very high depth. In order to significantly increase the depth over the previous models~\cite{Krizhevsky-NIPS-2012,Simonyan-ICLR-2014}, He et al.~\cite{He-CVPR-2016} introduced skip connections to eliminate vanishing gradient problems in training very deep neural models, providing alternative pathways for the gradients during back-propagation. We opted for different residual architectures for each data set, adjusting each architecture and ensemble to match the complexity of the task and the number of data samples. Depending on the data set, our ensemble is typically formed of four or five residual nets.

\noindent {\bf Self-paced ensemble learning.}
In self-paced learning, machine learning models learn at their own pace, taking into consideration the samples with high confidence predictions first. This technique is robust in solving problems with noisy labels \cite{meng2017theoretical}. In our work, we consider an unsupervised domain adaptation scenario, in which the ground-truth labels of the target domain samples used for SPL are unknown. Hence, the model needs to learn from its own labels, called pseudo-labels. We propose an automatic self-paced ensemble learning technique that iteratively enriches the training data with self-annotated examples from the target domain. The proposed approach assumes no overhead at testing time, the network architectures being identical (we only update the models' weights). The steps required by our framework are formally described in Algorithm \ref{alg_1}. The algorithm is divided into three stages. The first two stages are about training the ensemble, while the third stage is about evaluating the ensemble on the final test set.

\begin{algorithm}[!t]
\caption{Self-paced ensemble learning (SPEL)}\label{alg_1}
\textbf{Input: }{$(X,T)$ - a training set of samples and associated labels from the source domain; $Y$ - a test set from the target domain; $\bar{Y}$ - an unlabeled set from the target domain; $n$ - the number of models in the ensemble; $k$ - the number of SPEL steps; $m$ - the number of extra examples added at a SPEL step; $\eta$ - a learning rate; $\mathcal{L}$ - a loss function.}\\
\textbf{Notations: }{$f_i$ - the $i$-th model in the ensemble; $\theta_i$ - the weights of the $i$-th model; \emph{avg} - a function that computes Bayesian model averaging; \emph{sort} - a function that jointly sorts the input set; $\mathcal{N}(0, \Sigma)$ - the normal distribution of mean $0$ and standard deviation $\Sigma$; $\mathcal{U}(S)$ - the uniform distribution over the set $S$; $S[i:j]$ - subset of elements $\{s_i,s_{i+1},...,s_j\}$ from $S$.}\\
\textbf{Initialization: }{$\theta^{(0)}_{i} \sim \mathcal{N}(0, \Sigma), \forall i \in \{1,...,n\}$}\\
\textbf{Output: }{$P_Y$ - the predictions for the test set $Y$.}
\vspace{0.2em}
\algrule
\vspace{0.2em}
\textbf{Stage 1:}{ Pre-training of individual models}
\vspace{0.2em}
\algrule
\begin{algorithmic}[1]
\For{$i \gets 1$ to $n$}
    \State $t \gets 0$
    \While{converge criterion not met}
     \State $X^{(t)}, T^{(t)} \gets$ mini-batch $\sim \mathcal{U}((X,T))$
     \State $\theta^{(t+1)}_{i} = \theta^{(t)}_{i} - \eta^{(t)} \nabla{\mathcal{L}\left(\theta^{(t)}_{i}, X^{(t)}, T^{(t)}\right)}$
     \State $t \gets t + 1$
    \EndWhile
\EndFor
\vspace{0.2em}
\algrule
\vspace{0.2em}
\noindent\hspace{-2em}
\textbf{Stage 2:}{ Self-paced ensemble learning}
\vspace{0.2em}
\algrule
\vspace{0.2em}
\For{$j \gets 1$ to $k$}
    \State $P_{\bar{Y}}, C_{\bar{Y}} \gets \mbox{\emph{avg}}\!\left( f_1\!\left(\theta^{(t)}_1, \bar{Y}\right),...,f_n\!\left(\theta^{(t)}_n, \bar{Y}\right)\!\right)$
    \State $\bar{Y}',P_{\bar{Y}'} \gets \mbox{\emph{sort}}\!\left(\bar{Y}, P_{\bar{Y}}\right)$ with respect to $C_{\bar{Y}}$
    \State $\bar{X}^{(j)} \gets \bar{Y}'[1:m \cdot j]$
    \State $\bar{T}^{(j)} \gets P_{\bar{Y}'}[1:m \cdot j]$
    \State $X' \gets X \cup \bar{X}^{(j)}$
    \State $T' \gets T \cup \bar{T}^{(j)}$
    \For{$i \gets 1$ to $n$}
    \While{converge criterion not met}
     \State $X^{(t)}, T^{(t)} \gets$ mini-batch $\sim \mathcal{U}((X',T'))$
     \State $\theta^{(t+1)}_{i} = \theta^{(t)}_{i} - \eta^{(t)} \nabla{\mathcal{L}\left(\theta^{(t)}_{i}, X^{(t)}, T^{(t)}\right)}$
     \State $t \gets t + 1$
    \EndWhile
\EndFor
\EndFor
\vspace{0.2em}
\algrule
\vspace{0.2em}
\noindent\hspace{-2em}
\textbf{Stage 3:}{ Prediction}
\vspace{0.2em}
\algrule
\vspace{0.2em}
\State $P_{Y} \gets \mbox{\emph{avg}}\!\left( f_1\!\left(\theta^{(t)}_1, {Y}\right),...,f_n\!\left(\theta^{(t)}_n, {Y}\right)\!\right)$
\end{algorithmic}
\end{algorithm}


In the first stage (steps 1-6), we employ the conventional training procedure based on stochastic gradient descent with mini-batches. Each neural network $f_i$ which is part of the ensemble is optimized until we reach an optimal convergence point. At the current iteration $t$, the first step (step 4) is to randomly sample a mini-batch of labeled examples from the training data set $(X,T)$. Then, at step 5, the weights $\theta_i$ corresponding to the network $f_i$ are updated in the negative direction of the gradient of the loss function $\nabla \mathcal{L}$, the update step being controlled through the learning rate $\eta$.

In the second stage (steps 7-18), we employ our self-paced ensemble learning procedure for a number of $k$ iterations, where $k$ is a pre-established hyperparameter of our method. At each iteration, we start by computing the predicted labels $P_{\bar{Y}}$ and the confidence scores $C_{\bar{Y}}$ provided by the ensemble (step 8) for the unlabeled set $\bar{Y}$, which belongs to the target domain. Next, at step 9, we sort the set of samples $\bar{Y}$ and the associated pseudo-labels $P_{\bar{Y}}$ with respect to $C_{\bar{Y}}$, in decreasing order of the confidence scores. Afterwards, we select $m\cdot j$ samples (step 10) and the corresponding pseudo-labels (step 11), which are added (steps 12 and 13) to the new training set $(X',T')$. We underline that at each iteration $j$, the number of extra samples is $m$, where $m$ is another pre-established hyperparameter of our method. As the confidence scores of the previously included samples may change over time, we re-evaluate them at each iteration $j$, regenerating $X'$ and $T'$. Hence, at iteration $j$, we add $m\cdot j$ pseudo-labeled samples. Upon forming the new training set $(X',T')$, we continue the training of the individual models, applying stochastic gradient descent with mini-batches in steps 14-18. We underline that, by eliminating the second training stage, we fall back on the baseline ensemble learning.

In the third and final stage (step 19), we apply our ensemble based on Bayesian model averaging on the test set $Y$, obtaining the final predictions $P_Y$. We note that the proposed self-paced ensemble learning algorithm can be applied on top of any type of ensemble and it does not involve any additional overhead at test time.

\section{Experiments}

\vspace{-0.1cm}
\subsection{Data sets}
\vspace{-0.1cm}

{\bf CREMA-D.}
The CREMA-D multi-modal database \cite{Cao-TAC-2014} contains 7,442 videoclips of 91 actors (48 male and 43 female) with different ethnic backgrounds. The actors were asked to simulate particular emotions while producing, with different intonations, 12 particular sentences that evoke the target emotions. There are six emotion categories: neutral, happy, anger, disgust, fear and sad. In our experiments, we consider only the audio modality as source of information. We split the audio samples into $70\%$ for training, $15\%$ for validation and $15\%$ for testing.

\noindent {\bf RCSAD.}
The Rainforest Connection Species Audio Detection (RCSAD) data set \cite{RFCx} is released by Rainforest Connection (RFCx) and is based on acoustic data collected while monitoring the ecosystem soundscape at tropical locations. RCSAD contains 4,727 audio files for training and 1,992 audio files for testing. We kept $20\%$ of the training set for validation.
The data set contains 24 classes representing bird and frog species from tropical regions. An audio sample may contain one or more species. Therefore, the goal is to solve a multi-label classification task. The data set is released in a recent Kaggle competition, implying that the only way to evaluate a model is by making a submission on Kaggle, i.e.~the test labels are private.

\noindent {\bf MSC.}
The Mask Augsburg Speech Corpus (MSC) is provided by the ComParE organizers \cite{Schuller-INTERSPEECH-2020}. It comprises recordings of 32 German native speakers, with or without wearing surgical masks. Each data sample is a recording of $1$ second at a sampling rate of $16$ KHz. The data set is partitioned into a training set of 10,895 samples, a development set of 14,647 samples and a test set of 11,012 samples. Since the test labels are private, we report results on the development set, keeping $20\%$ of the training set for validation.

\vspace{-0.1cm}
\subsection{Evaluation setup}
\vspace{-0.1cm}

\noindent {\bf Performance measures.} 
For the CREMA-D data set, we report the classification accuracy. For the RCSAD data set, the Kaggle competition organizers decided to rank participants based on the weighted label-ranking average precision (WLRAP). Therefore, we report our performance in terms of this measure. Regarding the MSC data set, we report the unweighted average recall (UAR), this being the official measure in the ComParE challenge \cite{Schuller-INTERSPEECH-2020}. 

\noindent {\bf Baselines.}
On all data sets, we consider as baseline the model ensemble trained without the second stage described in Algorithm~\ref{alg_1}. For the CREMA-D data set, we added as baselines two state-of-the-art methods \cite{Shukla-ICASSP-2020, He-CVPRW-2020} that reported the accuracy on the audio modality. For RCSAD, we refer to our ranking in the Kaggle competition, as there are no prior publications reporting results on this data set. For both CREMA-D and RCSAD, we also report the results of the best model in the baseline ensemble. For the MSC data set, we considered the official competition baseline as well as the approach proposed in \cite{ristea2020you}, the latter approach coinciding with our baseline model ensemble.

\noindent {\bf Data preprocessing and tuning.}
We first standardized all audio clips to a fixed dimension. For CREMA-D, we padded or clipped the audio files to 4 seconds, while for RCSAD, we extracted an 8 second window centered on each training detection and we applied the sliding window algorithm at test time, keeping the maximum probabilities. The length of the MSC audio files is already established by the organizers at 1 second. When applying the STFT, we used $N_{x}\!=\!1024$, $R\!=\!64$ and a window size of $512$. Each utterance is represented as a Mel spectrogram of $1\!\times\!L\!\times\!256$ components, where $256$ is the number of Mel bins and $L$ represents the number of time bins, which is $512$ for CREMA-D and $618$ for RCSAD. Exceptionally, on the MSC data set, we performed the preprocessing proposed in \cite{ristea2020you}, which involves only STFT.

While our ensembles are all based on residual nets, we used slight different architectures on the three data sets. For CREMA-D, we employed an ensemble of five ResNet-18 models \cite{He-CVPR-2016} with leaky ADA activations \cite{georgescu2020non} which increase the model's capability to classify non-linearly separable data, such as emotions from voice recordings. For RCSAD, we considered an ensemble of five ResNeSt-50 models \cite{zhang2020resnest}. We modified the classification layer of the ResNeSt-50 network to predict 24 classes and added a sigmoid activation function, such that the network is able to output the probability of each class independently of the other classes. For MSC, the ensemble comprises a ResNet-18, a ResNet-34, a ResNet-50 and a ResNet-101, being identical to the ensemble proposed by Ristea et al.~\cite{ristea2020you}.
 
All models are optimized with Adam \cite{Kingma-ICLR-2015}. For CREMA-D, we trained the ResNet-18 models for $70$ epochs on mini-batches of $16$ samples using a learning rate of $5 \cdot 10^{-4}$. For RCSAD, we set the learning rate to $10^{-3}$ and trained the models for $30$ epochs on mini-batches of $16$ samples. For MSC, we used the same hyperparameters as Ristea et al.~\cite{ristea2020you}. Our SPEL framework, has two additional hyperparameters. For the parameter $m$, we considered values in the set $\{50,100,150,200\}$. For the parameter $k$, we considered all possible values subject to $m \cdot k \leq 1,000$, i.e.~we added at most $1,000$ pseudo-labeled samples during our second training stage. On CREMA-D, we obtained optimal results with $m=50$ and $k=3$. On the other two data sets, we obtained optimal results with $m=150$ and $k=4$. All hyperparameters are tuned on the validation sets.

\begin{table}[!t]
\caption{Results of our SPEL approach versus various baseline and state-of-the-art methods on CREMA-D. Significantly better self-paced learning results are marked with $\ddagger$, using a paired McNemar's test \cite{Dietterich-NC-1998} at the significance level $0.01$.}
\label{tab_results_cremad}
\vspace{-0.2cm}
\centering
\begin{tabular}{l c}
\toprule
{\textbf{Method}} & {\textbf{Accuracy}}\\
\midrule
Shukla et al.~\cite{Shukla-ICASSP-2020} & 55.01 \\
He et al.~\cite{He-CVPRW-2020} & 58.71 \\
\midrule
ResNet-18 + leaky ADA \cite{georgescu2020non}          & 65.37 \\
ResNet-18 + leaky ADA \cite{georgescu2020non} + SPL          & 66.04$^\ddagger$ \\
\midrule
$5\times$ResNet-18 + leaky ADA \cite{georgescu2020non}          & 66.54 \\
$5\times$ResNet-18 + leaky ADA \cite{georgescu2020non} + SPEL   & 68.12$^\ddagger$ \\
\bottomrule
\end{tabular}
\vspace{-0.1cm}
\end{table}

\begin{table}[!t]
\caption{Results of our SPEL approach versus several baselines on RCSAD. Significantly better self-paced learning results are marked with $\ddagger$, using a paired McNemar's test \cite{Dietterich-NC-1998} at the significance level $0.01$.}
\label{tab_results_rcsad}
\centering
\vspace{-0.2cm}
\begin{tabular}{l c c}
\toprule
\textbf{Method} & \textbf{WLRAP} & \textbf{Kaggle Rank}\\
\midrule
ResNeSt-50                              & 0.862 & 521 / 1143\\
\midrule
$5\times$ResNeSt-50                     & 0.900 & 98 / 1143\\
$5\times$ResNeSt-50 + SPEL               & {0.906}$^\ddagger$ & 84 / 1143\\
$5\times$ResNeSt-50 + SPEL + PP          & {0.943}$^\ddagger$ & 28 / 1143\\
\bottomrule
\end{tabular}
\vspace{-0.4cm}
\end{table}

\vspace{-0.cm}
\subsection{Results}
\vspace{-0.1cm}

In Tables \ref{tab_results_cremad}, \ref{tab_results_rcsad} and \ref{tab_results_msc}, we present the results obtained on three different data sets by our SPEL approach against various baseline methods. We observe that SPEL attains the highest improvements on CREMA-D (see Table~\ref{tab_results_cremad}), where it outperforms all state-of-the-art results with an accuracy of $68.12\%$. The improvement of  $1.58\%$ over the corresponding baseline ensemble is statistically significant according to a paired McNemar's test \cite{Dietterich-NC-1998} performed at the significance level $0.01$. Additionally, we empirically demonstrate that our SPEL approach brings a higher gain (of $1.58\%$) than the gain (of $0.67\%$) of SPL applied on top of a single model. This indicates that the models in the ensemble are able to boost each others' accuracy.

\begin{table}[!t]
  \caption{Results of our SPEL approach versus the official competition baseline and the ensemble proposed by Ristea et al.~\cite{ristea2020you} on MSC. Significantly better self-paced learning results are marked with $\ddagger$, using a paired McNemar's test \cite{Dietterich-NC-1998} at the significance level $0.01$.}
  \label{tab_results_msc}
  \centering
\vspace{-0.2cm}
\begin{tabular}{l c}
\toprule
{\textbf{Method}} & {\textbf{UAR}}\\
\midrule
DeepSpectrum \cite{Schuller-INTERSPEECH-2020}                   & 63.4 \\
Ristea et al. \cite{ristea2020you}                              & 72.2 \\
\midrule
Ristea et al. \cite{ristea2020you} + SPEL                        & {72.5}$^\ddagger$ \\
\bottomrule
\end{tabular}
\vspace{-0.1cm}
\end{table}

\begin{figure}[t!]
\centering
\includegraphics[width=1.0\linewidth]{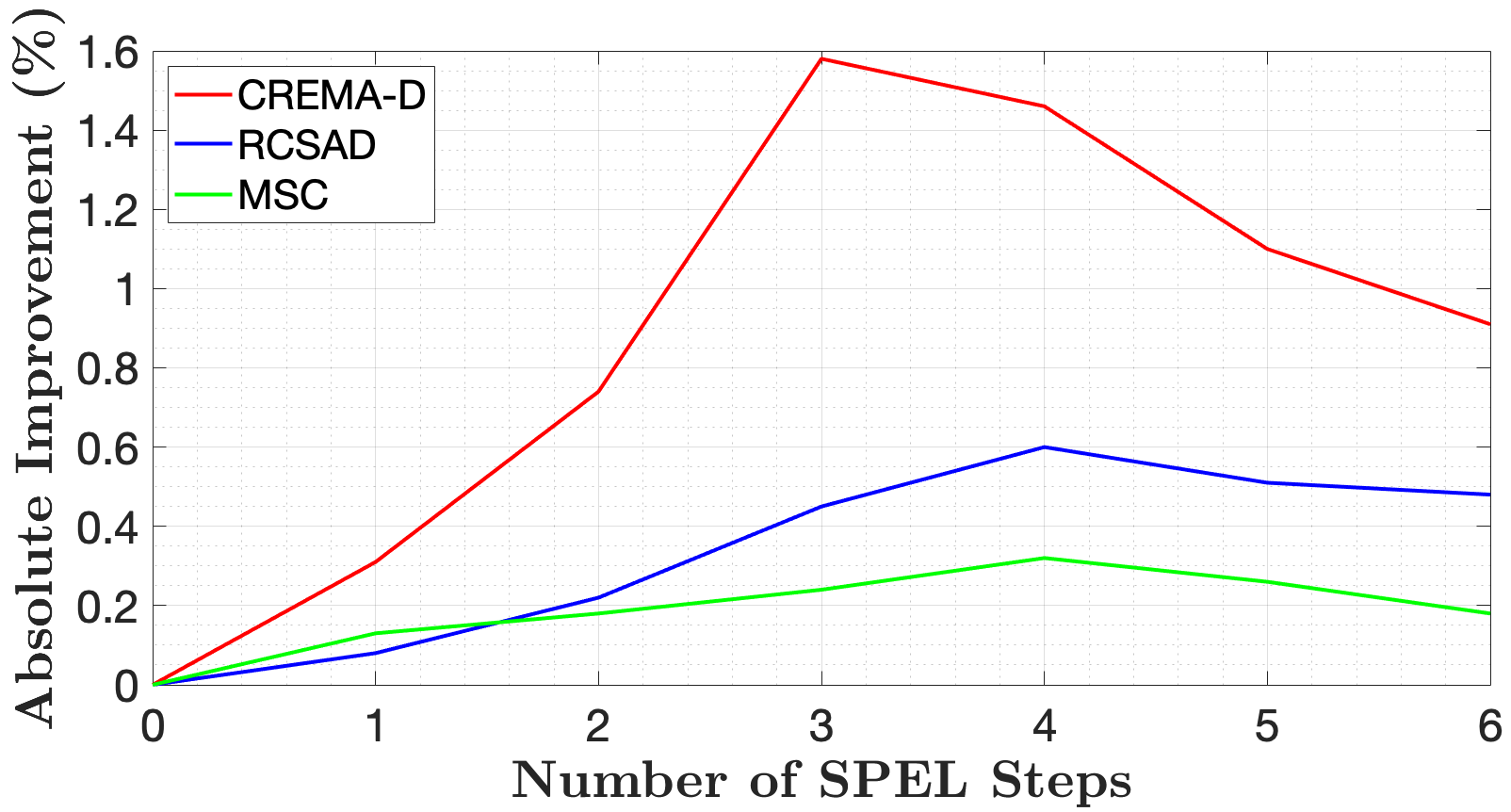}
  \vspace{-0.5cm}
  \caption{The absolute improvements brought by our SPEL algorithm at various steps, with respect to the corresponding baseline model ensemble. Best viewed in color.}
  \label{fig_rel_imp}
  \vspace{-0.3cm}
\end{figure}

On RCSAD (see Table~\ref{tab_results_rcsad}), the ensemble model trained with SPEL surpasses the baseline ensemble by a statistically significant margin of $0.6\%$. We obtained an additional post-competition performance boost (of $3.7\%$), when we learned that a cheap post-processing (PP)\footnote{\url{https://www.kaggle.com/c/rfcx-species-audio-detection/discussion/220389}} of the probabilities is useful.

In Table~\ref{tab_results_msc}, we observe that the model proposed by Ristea et al.~\cite{ristea2020you} gains a performance boost of $0.3\%$ when SPEL is applied. Even if the magnitude of the performance gain seems small, a paired McNemar's test \cite{Dietterich-NC-1998} performed at the significance level $0.01$ reveals that the difference is significant.

In Figure~\ref{fig_rel_imp}, we show the absolute performance gains brought by our approach with respect to the number of SPEL steps $k$. We observe that the generic behavior of SPEL is to increase the performance until a certain point (typically, after $3$ to $5$ steps), and afterwards, the performance slowly declines. We note that SPEL brings in pseudo-labeled examples during training, which means that a small percentage of the newly added sample have incorrect labels. The proportion of incorrectly labeled samples increases with the number of SPEL steps. Hence, when the signal-to-noise ratio of the labels becomes unfavorable, the performance starts to decline. 

\section{Conclusion}

In this paper, we presented a self-paced ensemble learning algorithm in which models learn from each other over several iterations. To demonstrate the generality of our SPEL scheme, we conducted comprehensive experiments with neural architectures of various depths on three audio tasks, where we obtained statistically significant performance gains. In future work, we aim to study increasing the regularization when the ensemble's performance starts to decline.

\noindent
{\bf Acknowledgements.} Work supported by a grant of the Romanian Ministry of Education and Research, CNCS - UEFISCDI, project no. PN-III-P1-1.1-TE-2019-0235, within PNCDI III.

\bibliographystyle{IEEEtran}
\bibliography{mybib}

\end{document}